\documentclass[aps,a4paper,preprint]{revtex4}

\usepackage{graphicx}
\usepackage{amssymb}

\begin{document}

\title{Stochastic Simulation of Nonadiabatic Dynamics at Long Time
}

\author{Daniel A. Uken and Alessandro Sergi}

\affiliation{
School of Physics, University of KwaZulu-Natal, Pietermaritzburg Campus,
Private Bag X01 Scottsville, 3209 Pietermaritzburg, South Africa }

\author{Francesco Petruccione}

\affiliation{
School of Physics and National Institute for Theoretical Physics,
University of KwaZulu-Natal, Westville Campus, Private Bag X54001,
Durban 4000, South Africa }

\begin{abstract}
Using a generalized transition probability,
it is shown how nonadiabatic calculations, within the Wigner-Heisenberg representation
of quantum mechanics, can be reliably extended to far longer times than
those allowed by a primitive sampling scheme.
Tackling the spin-boson model as a paradigmatic example,
substantial numerical evidence is provided that 
the statistical error can be reduced for a wide range of 
parameter values.
\end{abstract}

\today

\maketitle

\section{Introduction}

Many quantum processes can be modelled by the time evolution of a relevant subsystem 
coupled
to a bath. In certain representations, the dynamics of the subsystem can be understood
in terms of the transitions between quantum states as a result of the nonadiabatic
interaction with the bath.
Various methods have been proposed to formulate numerical algorithms
for nonadiabatic dynamics~\cite{sh-orig,heller,shenvi,book-qc}.
The main theoretical difficulty is associated with a rigorous description
of the back-reaction~\cite{qc} of the subsystem on the bath (or a systematic way
to derive controllable approximations of this phenomenon).

A systematic approach for deriving approximations of nonadiabatic quantum dynamics
has been proposed some time ago~\cite{qc}:
One can start from the Heisenberg equations of motion of the system
and perform
a partial Wigner transform~\cite{wigner} over only the degrees of freedom of the bath.
In such a way, one obtains what may be called Wigner-Heisenberg representation
of quantum mechanics~\cite{aif}. 
For harmonic baths, such a Wigner-Heisenberg representation
leads to an exact formulation of quantum dynamics.
For more general baths, a linear approximation of the propagator
leads to a hybrid quantum-classical representation~\cite{kc}.
In this case, time evolution is dictated by what is known as 
quantum-classical Liouville equation~\cite{qcl}. This equation
generates a unitary dynamics for all the degrees of freedom in the system.
Integrating over the phase space degrees of freedom~\cite{dissipative-qcl},
one can obtain stochastic equations for the remaining degrees of freedom
satisfying the quantitative criteria analyzed in~\cite{kohen}.
One of the appealing features
of the Wigner-Heisenberg formulation of mixed quantum-classical systems
is that it leads
to a rigorous formulation of their statistical mechanics~\cite{nielsen}.
In both cases, controllable approximations lead to a consistent
formulation of nonadiabatic dynamics, which is energy-conserving 
and correctly describes the back-reaction mentioned above.
From the computational point of view, this requires one
to evaluate statistically the propagator in terms
of a stochastic process describing the occurrence of quantum transitions:
Wigner-Heisenberg quantum mechanics can be represented
in terms of piecewise adiabatic trajectories of the bath coordinates,
interspersed by stochastic transitions between the energy levels
of the subsystem.
Despite the elegance of such an approach,
the growth in time of the amplitude of the statistical fluctuations
has so far somewhat limited the length of the time interval
that can be explored. 
In order to tame this problem two approaches must be combined.
The first is to develop more accurate approximations for the short-time
expression of the Wigner-Heisenberg propagator~\cite{trotter}.
This is critical especially for high values of the friction.
The other is to improve the time-dependent stochastic sampling 
of nonadiabatic
quantum transitions~\cite{af}.
Here we are concerned with this second issue.

In the following, starting from a simple approximation
of the short-time Wigner-Heisenberg propagator,
we discuss in detail how a sampling scheme recently 
introduced in~\cite{af} can  dramatically improve
the numerical stability of the calculations at long time 
(depending upon the numerical parameters,
the time span can be up to an order of magnitude longer 
than in previous schemes).
Our sampling method is based upon a suitable generalization of the
transition probability that effectively filters 
some trajectories which would cause an excessive energy fluctuation, 
within an approximate representation
of the back-reaction. Such a filtering
limits the growth in time of the statistical fluctuations.
This in turn allows the numerical method to access 
far longer integration times.

The structure of this paper is the following.
In Sec.~\ref{sec:wh} we briefly sketch the theory of the
Wigner-Heisenberg representation of quantum mechanics.
Section~\ref{sec:mj} illustrates how nonadiabatic dynamics
can be derived within the Momentum-Jump (MJ) approximation.
In Section~\ref{sec:w} we introduce a generalized sampling
scheme for the nonadiabatic propagator.
In Section~\ref{sec:sb} we present the results of some numerical calculations
on the nonequilibrium dynamics of the spin-boson model.
Finally, in Sec.~\ref{sec:c} we discuss
our achievements and propose future directions.


\section{Wigner-Heisenberg Representation of Quantum Mechanics}
\label{sec:wh}

Consider a system defined by the total Hamiltonian operator
\begin{equation}
\hat{H}=\hat{H}_S+\hat{H}_B+\hat{H}_{SB}\;,
\label{eq:tot-ham}
\end{equation}
where the subscripts $S$, $B$ and $SB$ stand for subsystem, bath
and coupling, respectively.
The Heisenberg equation of motion of a generic operator $\hat{\chi}$
can be written as~\cite{b3}
\begin{eqnarray}
\frac{\partial}{\partial t}\hat{\chi}
&=&\frac{i}{\hbar}\left[\begin{array}{cc} \hat{H} & 
\hat{\chi}\end{array}\right]
\cdot\mbox{\boldmath$\cal B$}^c\cdot
\left[\begin{array}{c}\hat{H}\\ \hat{\chi}\end{array}\right]
\;,\label{eq:heisenberg}
\end{eqnarray}
where $\mbox{\boldmath$\cal B$}^c$ is the symplectic matrix~\cite{goldstein}
\begin{equation}
\mbox{\boldmath$\cal B$}^c=\left[\begin{array}{cc} 0 & 1 \\ -1 & 0\end{array}\right]
\;.
\end{equation}

We assume that the bath Hamiltonian depends upon a pair of canonically 
conjugate operators, $\hat{X}=(\hat{R},\hat{P})$, and that the coupling has
the form $\hat{H}_{SB}=\hat{H}_{SB}(\hat{R})$.
A partial Wigner transform 
can be introduced as
\begin{eqnarray}
\hat{\chi}_W(X)
&=&
\int dz e^{iP\cdot z/\hbar}
\langle R-\frac{z}{2}\vert\hat{\chi}
\vert R+\frac{z}{2}\rangle \;,
\end{eqnarray}
for the generic bath-dependent operator $\hat{\chi}$
and, analogously, for the density matrix $\hat{\rho}$ as
\begin{eqnarray}
\hat{\rho}_W(X)
&=&\frac{1}{(2\pi\hbar)^{3N}}
\int dz e^{iP\cdot z/\hbar}
\langle R-\frac{z}{2}\vert\hat{\rho}
\vert R+\frac{z}{2}\rangle \;,
\end{eqnarray}
where $X=(R,P)$ are canonically conjugate classical 
variables in phase space.
The Wigner-Heisenberg equations of motion
are obtained 
upon taking the partial Wigner transform of Eq.~(\ref{eq:heisenberg}):
\begin{eqnarray}
\frac{\partial}{\partial t}\hat{\chi}_W(X,t)
&=&\frac{i}{\hbar}\left[\begin{array}{cc} \hat{H}_W(X) & 
\hat{\chi}_W\end{array}\right]
\cdot\mbox{\boldmath$\cal D$}\cdot
\left[\begin{array}{c}\hat{H}_W(X)\\ \hat{\chi}_W(X,t)\end{array}\right]
\;,\nonumber\\
\label{eq:pW}
\end{eqnarray}
where
\begin{eqnarray}
\mbox{\boldmath$\cal D$}
&=&
\left[\begin{array}{cc} 0 & e^{\frac{i\hbar}{2}
\overleftarrow{\partial}_k{\cal B}_{kj}^c
\overrightarrow{\partial}_j}
\\
-e^{\frac{i\hbar}{2}
\overleftarrow{\partial}_k{\cal B}_{kj}^c
\overrightarrow{\partial}_j}
& 0\end{array}\right]\;.\label{eq:defD}
\end{eqnarray}
In Eq.~(\ref{eq:defD}) we have used the symbols
$\overrightarrow{\partial}_k=\overrightarrow{\partial}/\partial X_k$
and
$\overleftarrow{\partial}_k=\overleftarrow{\partial}/\partial X_k$
to denote the operators of derivation (with respect to the phase space point
coordinates) acting on the right or left, depending on the direction of
the arrow.  The sum over repeated indices must be understood in
Eq.~(\ref{eq:defD}) and in the following.
The mixed Wigner-Heisenberg form of the Hamiltonian
operator, $\hat{H}_W$, is 
\begin{equation}
\hat{H}_W(X)=\hat{H}_S+H_{W,B}(X)+\hat{H}_{W,SB}(R)\;.
\end{equation}

Equation~(\ref{eq:pW}) provides a mixed Wigner-Heisenberg
representation of quantum mechanics, where operators also depend
upon phase space (c-number) coordinates.
Such a representation is completely equivalent
to the usual Heisenberg representation, but
calculations are very difficult in general.
However, in the case of quadratic bath Hamiltonians
\begin{equation}
\hat{H}_{W,B}=\sum_{k=1}^N\left
(\frac{P_k^2}{2}+\frac{1}{2}\omega_k^2R_k^2\right)\;,
\label{eq:hambqua}
\end{equation}
where $(R_k,P_k)$, $k=1,\ldots,N$, are the coordinates and momenta,
respectively,
of a system of $N$ independent harmonic oscillators with frequencies
$\omega_k$,
and for interaction Hamiltonians of the type
\begin{equation}
\hat{H}_{W,SB}=V_B(R)\otimes \hat{H}_{S}^{\prime}\;,
\label{eq:hamcqua}
\end{equation}
where $V_B(R)$ is at most a quadratic function of $R$
and $\hat{H}_S^{\prime}$ acts only in the Hilbert space
of the subsystem,
Eq.~(\ref{eq:pW}) can be rewritten using
the antisymmetric operator matrix
\begin{eqnarray}
\mbox{\boldmath$\cal D$}_{\rm lin}
&=&
\left[\begin{array}{cc} 0 & 1+\frac{i\hbar}{2}
\overleftarrow{\partial}_k{\cal B}_{kj}
\overrightarrow{\partial}_j
\\
-1-\frac{i\hbar}{2}
\overleftarrow{\partial}_k{\cal B}_{kj}
\overrightarrow{\partial}_j
& 0\end{array}\right]\;.
\label{eq:dlin}
\nonumber\\
\end{eqnarray}
Actually, it can be shown that for the class of Hamiltonians
specified by Eqs.~(\ref{eq:hambqua}) and~(\ref{eq:hamcqua})
\begin{equation}
\mbox{\boldmath$\cal D$}\to\mbox{\boldmath$\cal D$}_{\rm lin}
\label{eq:Dlinsub}
\end{equation}
holds exactly.
For more general bath Hamiltonians and couplings, such a substitution
amounts to performing a quantum-classical approximation~\cite{kc}.
What matters here is that for the class of systems in which we are interested,
Eq.~(\ref{eq:Dlinsub}) is exact and provides 
via Eq.~(\ref{eq:heisenberg}) a Wigner-Heisenberg formulation of quantum
mechanics which can be numerically simulated~\cite{theorchemacc}.


\section{Momentum-Jump Nonadiabatic dynamics}
\label{sec:mj}

In order to devise a numerical algorithm, Eq.~(\ref{eq:pW})
must be represented in some basis.
The adiabatic basis, defined by
\begin{equation}
\hat{H}_W(X)=\frac{P^2}{2M}+\hat{h}_W(R)\;,
\end{equation}
and the eigenvalue equation,
\begin{equation}
\hat{h}_W(R)|\alpha;R\rangle=E_{\alpha}(R)|\alpha;R\rangle\;,
\end{equation}
naturally leads to surface-hopping schemes, as is clarified below.
In such a basis the quantum-classical evolution is
\begin{equation}
\chi_W^{\alpha\alpha'}(X,t)
=\sum_{\beta\beta'}\left(e^{it{\cal L}}\right)_{\alpha\alpha',\beta\beta'}
\chi_W^{\beta\beta'}(X)\;,
\end{equation}
where
\begin{eqnarray}
i{\cal L}_{\alpha\alpha',\beta\beta'}
&=&\left(i\omega_{\alpha\alpha'}+iL_{\alpha\alpha'}\right)
\delta_{\alpha\beta}\delta_{\alpha'\beta'}
+J_{\alpha\alpha',\beta\beta'}
\\
&=&i{\cal L}^0_{\alpha\alpha'}
\delta_{\alpha\beta}\delta_{\alpha'\beta'}
+J_{\alpha\alpha',\beta\beta'}
\end{eqnarray}
is the quantum(-classical) Liouville operator~\cite{kc}.
In turn, the Liouville operator is defined in terms
of the Bohr frequency,
\begin{equation}
\omega_{\alpha\alpha'}(R)
=\frac{E_{\alpha}(R)-E_{\alpha'}(R)}{\hbar}\;,
\end{equation}
the classical-like Liouville operator for the bath degrees of freedom
\begin{equation}
iL_{\alpha\alpha'}=\frac{P}{M}\cdot\frac{\partial}{\partial R}
+\frac{1}{2}\left(F_W^{\alpha}+F_W^{\alpha'}\right)\cdot
\frac{\partial}{\partial P}\;,
\end{equation}
and the transition operator
\begin{eqnarray}
J_{\alpha\alpha',\beta\beta'}
&=&
\frac{P}{M}\cdot d_{\alpha\beta}(R)
\left(1+\frac{1}{2}
\frac{\Delta E_{\alpha\beta}(R)d_{\alpha\beta}(R)}
{\frac{P}{M}\cdot d_{\alpha\beta}(R)}\cdot\frac{\partial}{\partial P}
\right)
\delta_{\alpha'\beta'}
\nonumber\\
&+&
\frac{P}{M}\cdot d^*_{\alpha'\beta'}(R)
\left(1+\frac{1}{2}
\frac{\Delta E_{\alpha'\beta'}(R)d^*_{\alpha'\beta'}(R)}
{\frac{P}{M}\cdot d^*_{\alpha'\beta'}(R)}\cdot\frac{\partial}{\partial P}
\right)
\delta_{\alpha\beta}\;,
\label{eq:TO}
\end{eqnarray}
where $\Delta E_{\alpha\beta}(R)=E_{\alpha}(R)-E_{\beta}(R)$.
The $J$ operator defined by Eq.~(\ref{eq:TO})
is responsible for the nonadiabatic transitions between the energy
levels of the quantum subsystem as a result of the coupling to the bath.


It is in general difficult to devise numerical algorithms that implement the
exact form of the transition operator in Eq.~(\ref{eq:TO}).
The Momentum-Jump approximation consists in approximating the $J$ operator
as
\begin{eqnarray}
J_{\alpha\alpha',\beta\beta'}
\approx
J_{\alpha\alpha',\beta\beta'}^{\rm MJ}
&=&
{\cal T}_{\alpha\to\beta}^{\rm MJ}
\delta_{\alpha'\beta'}
+
{\cal T}_{\alpha'\to\beta'}^{*{\rm MJ}}
\delta_{\alpha\beta}\;,
\end{eqnarray}
where
\begin{equation}
{\cal T}_{\alpha\to\beta}^{\rm MJ}
=
\frac{P}{M}\cdot d_{\alpha\beta}(R)
\exp\left[\frac{1}{2}
\frac{\Delta E_{\alpha\beta}(R)d_{\alpha\beta}(R)}
{\frac{P}{M}\cdot d_{\alpha\beta}(R)}\cdot\frac{\partial}{\partial P}
\right]
\;.
\end{equation}
In this simplified form, the action of $J^{\rm MJ}$ on the momenta $P$ (the 
back-reaction)
when a quantum transition occurs can be calculated exactly.

For simplicity, in the following we consider that all masses $M$ are equal. 
When an $\alpha\to\beta$ transition occurs,
the application of the momentum-jump operator
to the momenta $P$ produces a shift,  $P\to P'$,
defined by
\begin{equation}
P'=P+\Delta_{\alpha\beta}^{\rm MJ}P
\;,
\end{equation}
where
\begin{eqnarray}
\Delta_{\alpha\beta}^{\rm MJ}P &=&
- 
\left(P\cdot \hat{d}_{\alpha\beta}\right) 
\hat{d}_{\alpha\beta}
\nonumber\\
&+&\hat{d}_{\alpha\beta}{\rm sign}(P\cdot \hat{d}_{\alpha\beta})
\sqrt{
\left(P\cdot \hat{d}_{\alpha\beta}
\right)^2 +M\Delta E_{\alpha\beta} }
\;,
\label{eq:ms-exact}
\end{eqnarray}
and $\hat{d}_{\alpha\beta}$ is the unit vector associated to the coupling vector
in the multidimensional space of all the particle coordinates.
The expansion of the square root on the right hand side of Eq.~(\ref{eq:ms-exact})
provides the approximated momentum-shift
\begin{equation}
\tilde{\Delta}_{\alpha\beta}^{\rm MJ}P
=\frac{1}{2}\frac{\Delta E_{\alpha\beta}(R)}{\frac{P}{M}\cdot \hat{d}_{\alpha\beta}}
\hat{d}_{\alpha\beta}\;.
\label{eq:ms-approx}
\end{equation}
The exact momentum-shift, given by Eq.~(\ref{eq:ms-exact}),
conserves the energy in every quantum transition.
Accordingly there are no problems with so-called ``frustrated hops''~\cite{truhlar}.
Instead, the property of energy conservation
is broken only by the approximated form of the momentum-shift
given by Eq.~(\ref{eq:ms-approx}).
In this work, the numerical propagation of the trajectory
is performed using the exact energy-conserving momentum-shift rule.
In the next section, we will discuss the sampling probability
for the quantum transitions.
Such a transition probability is not uniquely fixed.
Hence, we can modify it using a filtering scheme based on
the approximated momentum-jump.
We show that such a filtering leads to 
a substantial reduction of the statistical error in the calculations.

\section{Stochastic Sampling of Nonadiabatic Dynamics}
\label{sec:w}

The formulas of the previous section can be implemented through
an elegant and simple algorithm which is based on a sequential
time step expansion of the Dyson propagator~\cite{sstp}.
For a small time step $\Delta t$ the quantum-classical propagator
is approximated as
\begin{equation}
\left(e^{i\Delta t{\cal L}^{\rm MJ}}\right)_{\alpha\alpha',\beta\beta'}
\approx
e^{i\Delta t{\cal L}^0_{\alpha\alpha'}}
\left(\delta_{\alpha\beta}\delta_{\alpha'\beta'}
+\Delta t J^{\rm MJ}_{\alpha\alpha',\beta\beta'}\right)\;.
\label{eq:sstp}
\end{equation}
One can prove that the concatenation of short time steps
according to Eq.~(\ref{eq:sstp})
reproduces exactly the Dyson integral expansion
of the operator 
$\exp\left(i\Delta t{\cal L}\right)_{\alpha\alpha',\beta\beta'}$~\cite{sstp}.
However, from a computational point of view, it is impossible to
evaluate exactly the concatenation of the short time steps.
What one can do is to consider the sequential short time propagator
in Eq.~(\ref{eq:sstp}) as a stochastic operator
and sample the action of the transition operator $J$ according to
suitable transition probability.
Within such a stochastic evaluation of the Dyson expansion
of the propagator, 
the action of the operator 
$i{\cal L}^{\rm (MJ)}_{\alpha\alpha',\beta\beta'}$
is that of generating a piecewise adiabatic process 
for the momentum and coordinate variables, $P$ and $R$, respectively,
over energy surfaces identified by the state labels $\alpha$ and $\alpha'$.

The practical implementation of the stochastic propagator
requires the sampling of the transitions.
A primitive choice for the transition probability is given by:
\begin{eqnarray} 
{\cal P}_{\alpha\beta}^{(0)}(X,\Delta t)
=\frac{|\frac{P}{M}\cdot d_{\alpha\beta}(R)|\Delta t}
{1+|\frac{P}{M}\cdot d_{\alpha\beta}(R)|\Delta t}\;,
\label{eq:prob-jump0}
\end{eqnarray} 
which also determines the probability of not making any transition
in the time interval $\Delta t$ as
\begin{eqnarray} 
{\cal Q}_{\alpha\beta}^{(0)}(X,\Delta t)&=&1- {\cal P}_{\alpha\beta}^{(0)}
\nonumber\\
&=&\frac{1}
{1+|\frac{P}{M}\cdot d_{\alpha\beta}(R)|\Delta t}\;.
\label{eq:prob-nojump0}
\end{eqnarray} 

One can exploit a certain arbitrariness in the definition of
the transition probabilities, Eqs.~(\ref{eq:prob-jump0}) 
and~(\ref{eq:prob-nojump0}), in order to define a filtering
of the nonadiabatic transitions.
As we illustrate by some numerical calculations,
this reduces the statistical noise at long time.
To this end, we consider the variation of the energy,
as calculated from the approximated momentum shift 
in Eq.~(\ref{eq:ms-approx}).
After the quantum transition one has
\begin{equation}
{\cal E}_{\alpha\beta}=\frac{P^{\prime 2}}{2M}+E_{\alpha}(R)
-\left(\frac{P^2}{2M}+E_{\beta}(R)\right)
\end{equation}
where
\begin{equation}
P'=P+\Delta P_{\alpha\beta}
\end{equation}
are the new momenta arising from the quantum transition
$\alpha\to\beta$.
We define new transition probabilities as
\begin{eqnarray} 
{\cal P}_{\alpha\beta}(X,\Delta t)
=\frac{\Delta t|\frac{P}{M}\cdot d_{\alpha\beta}(R)|
\omega\left(c_{\cal E},{\cal E}_{\alpha\beta}\right) }
{1+\Delta t|\frac{P}{M}\cdot d_{\alpha\beta}(R)|
\omega\left(c_{\cal E},{\cal E}_{\alpha\beta}\right) }\;,
\label{eq:prob-jump}
\end{eqnarray} 
and
\begin{eqnarray} 
{\cal Q}_{\alpha\beta}(X,\Delta t)&=&1- {\cal P}_{\alpha\beta}
\nonumber\\
&=&\frac{1}
{1+\Delta t|\frac{P}{M}\cdot d_{\alpha\beta}(R)|
\omega\left(c_{\cal E},{\cal E}_{\alpha\beta}\right) }\;,
\label{eq:prob-nojump}
\end{eqnarray} 
where
\begin{equation}
\omega\left(c_{\cal E},{\cal E}_{\alpha\beta}\right)
=\left\{
\begin{array}{ccc}
 & 1& ~{\rm if} ~{\cal E}_{\alpha\beta}\le c_{\cal E} \\
 & 0 & ~{\rm otherwise}
\end{array}
\right.
\;.\label{eq:whnw}
\end{equation}
The numerical parameter $c_{\cal E}$ 
controls the amplitude of the energy fluctuations in a transition.

Eq.~(\ref{eq:prob-jump}) defines a generalization of 
the primitive sampling scheme.
The analytical form of the weight $\omega$ is to a large extent arbitrary. 
The choice adopted in the calculations reported in this paper
and in the previous~\cite{af} is devised by requiring that
the energy fluctuations caused 
by the approximated momentum-shift rule~(\ref{eq:ms-approx})
are not too large.
In what follows, we illustrate by means of numerical calculations
that the particular choice given by Eq.~(\ref{eq:whnw}) leads to 
a dramatic reduction of the statistical noise of the algorithm.

\section{Numerical Example}
\label{sec:sb}

In order to illustrate the effectiveness of our generalized sampling 
scheme, defined by Eqs.~(\ref{eq:prob-jump}-\ref{eq:whnw}),
we report the results of calculations performed on the 
nonequilibrium quantum dynamics of the spin-boson model.
Such a model is defined 
(using adimensional coordinates~\cite{theorchemacc}) by
\begin{eqnarray}
\hat{H}_S&=&-\Omega\hat{\sigma}_x
\\
\hat{H}_{W, SB}&=&-\hat{\sigma}_z\sum_{j=1}^Nc_jR_j\;,
\end{eqnarray}
where $\hat{\sigma}_x$ and $\hat{\sigma}_z$ are the spin Pauli matrices
and $c_j$ are numerical coefficients whose value is fixed by requiring
that the spectral density of the system has a Ohmic form.
The bath Hamiltonian is given by Eq.~(\ref{eq:hambqua}).
For this model the adiabatic basis is known analytically~\cite{leggett}.
One can think of situations in which at $t=0$ the spin is
excited in state up and the quantum harmonic modes are at thermal 
equilibrium as if there were no coupling before $t=0$.
For later times, the coupling can be switched on
and one can calculate, for example, the decay of the population,
$\langle\hat{\sigma}_z(t)\rangle$.
The dynamics of the spin-boson model has been well studied~\cite{leggett},
and thus provides us with a good system to check the efficacy of our
generalized sampling.

\begin{figure}
\includegraphics* {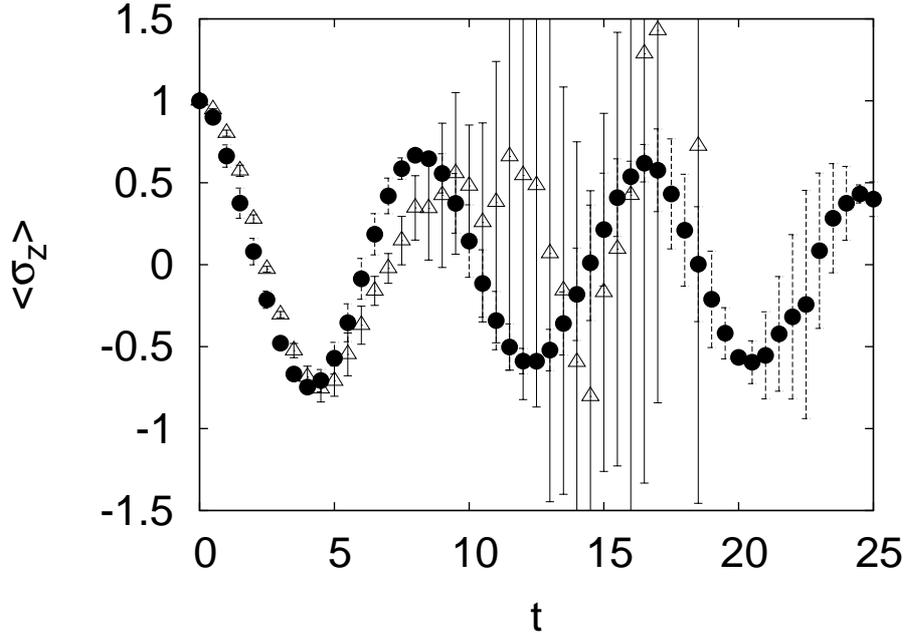}
\caption{
Comparison of primitive ($\triangle$) and generalized
($\bullet$) sampling
for $\beta=3$, $\Omega=1/3$, and $\xi=0.1$.
A parameter value of $c_{\cal E}=0.05$ was used.
Two quantum transitions per trajectory were included.
The error bars associated the primitive sampling start 
growing from $t=10$ and become enormous after such threshold.
The error bars associated to the generalized sampling become
distinguishable from the bullet only after $t=16$
and remain, nevertheless, very small up to $t=20$.
}
\label{fig:fig1}
\end{figure}

\begin{figure}
\includegraphics* {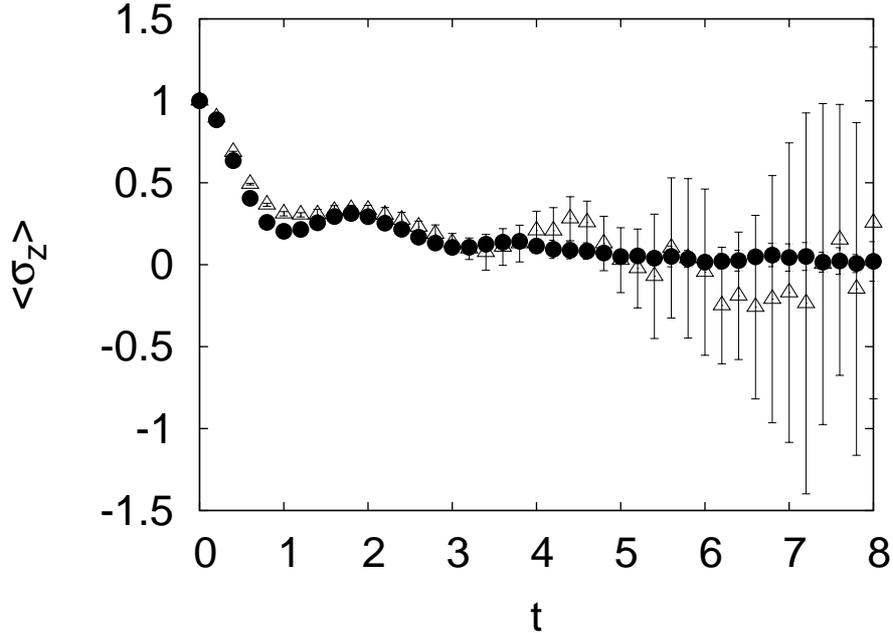}
\caption{
Comparison of primitive ($\triangle$) and generalized
($\bullet$) sampling
for $\beta=0.25$, $\Omega=1.2$, and $\xi=2$.
A parameter value of $c_{\cal E}=0.1$ was used.
Two quantum transitions per trajectory were included.
The picture shows clearly the error bars associated
with the primitive sampling. Those associated
with the generalized sampling are pretty much indistinguishable
from the bullet symbol up to $t=8$.
}
\label{fig:fig2}
\end{figure}

\begin{figure}
\includegraphics* {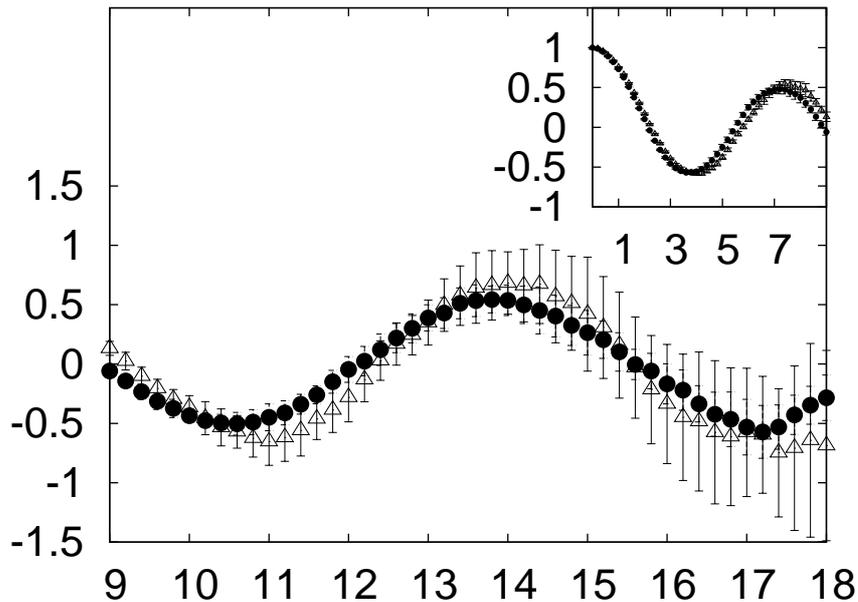}
\caption{
Comparison of primitive ($\triangle$) and generalized
($\bullet$) sampling
for $\beta=1$, $\Omega=0.4$, and $\xi=0.13$.
A parameter value of $c_{\cal E}=0.1$ was used.
Two quantum transitions per trajectory were included.
The inset shows the short-time behaviour where the two sampling schemes
provides very close numerical results.
The main figure shows the long-time behavior.
The error bars associated to the primitive sampling are clearly seen.
Those associated with the generalized sampling are barely distinguishable
from the bullet symbol only after $t=16$.
}
\label{fig:fig3}
\end{figure}

Here we compare the results of calculations performed
with the generalized sampling scheme to those of the
primitive algorithm.
In all calculations a maximum number of $n=2$ nonadiabatic transitions
per trajectory has been considered.
For each matrix element propagated in time,
the phase space ensemble 
has been composed of a number $N_{\rm mcs}=2.5\times 10^5$ points.
The integration time step of the phase space trajectory
has been taken as $dt=0.01$ in dimensionless units~\cite{theorchemacc}.
In Figs.~\ref{fig:fig1},~\ref{fig:fig2} and~\ref{fig:fig3} 
we display the results for the observable 
$\langle\hat{\sigma}_{z}\rangle$ against time for three sets of parameters.
Figure~\ref{fig:fig1} shows the results obtained
for the parameters $\beta=3$, $\Omega=1/3$ and $\xi=0.1$.
The generalized sampling with a parameter $c_{\epsilon}=0.05$
stabilizes the dynamics over a time-interval four to five times longer
than that achieved with the primitive sampling scheme.
Figure~\ref{fig:fig2} displays the results for 
$\beta=0.25$, $\Omega=1.2$ and $\xi=2$,
which represents higher friction for the spin-boson dynamics.
In this case, using $c_{\epsilon}=0.1$, the generalized sampling
allows one to extend the dynamics for at least a twice as long a time interval
than that of the primitive sampling.
Longer sampling is also achieved for $\beta=1$, $\Omega=0.4$ and $\xi=0.13$
with $c_{\epsilon}=0.1$.  Results are displayed in Fig.~\ref{fig:fig3}.

\begin{figure}
\includegraphics* {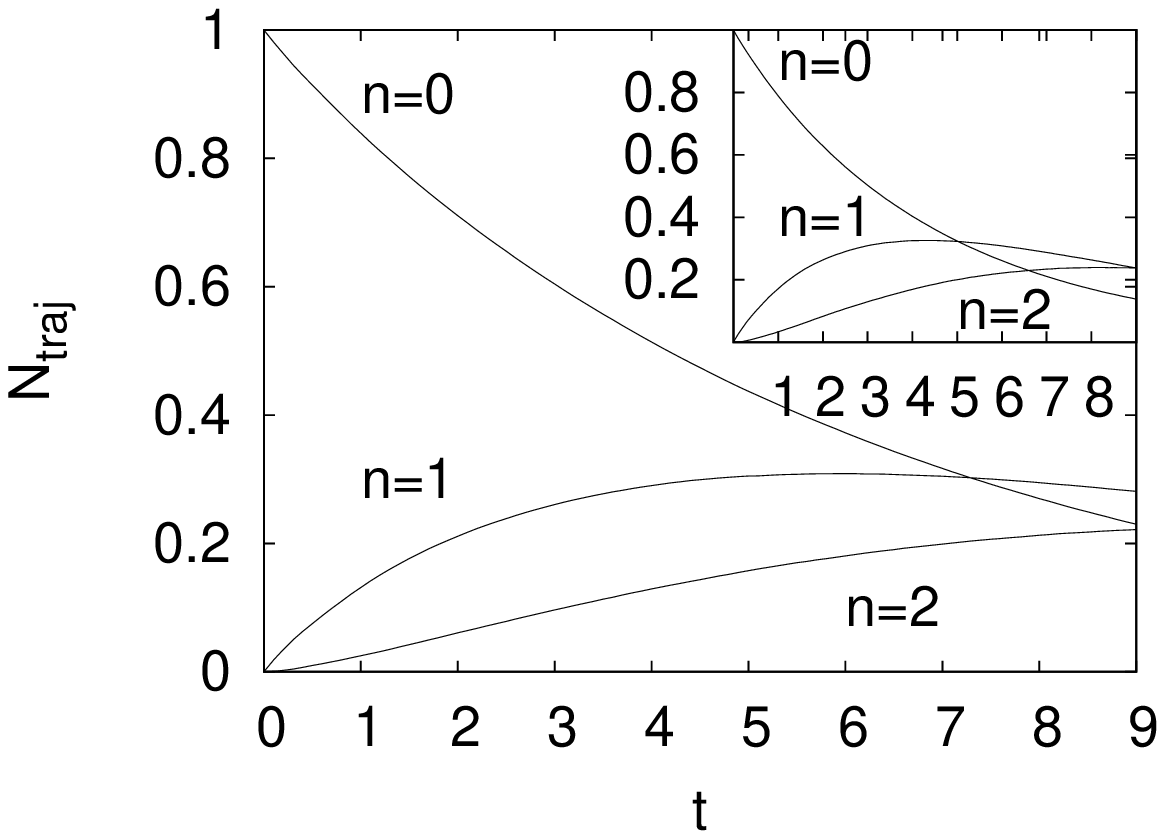}
\caption{
The fraction of trajectories in the ensemble involving $n=0$ (adiabatic dynamics
with no transitions) and $n=1,2$ nonadiabatic transition versus time.
The main figure shows the statistics in the case of the generalized sampling.
The inset displays the fraction of trajectories in the case of the primitive sampling.
The total ensemble was composed of $N=25\times 10^4$ trajectories in a calculation
with $\beta=1.0$, $\Omega=0.4$, and $\xi=0.13$.
}
\label{fig:jumpstat}
\end{figure}

The introduction of the weight $w$, defined in Eq.~(\ref{eq:whnw}),
in the generalized transition probability given by Eq.~(\ref{eq:prob-jump})
realizes a pruning in the ensemble of the sampled quantum transitions, while
keeping the important part of the nonadiabatic effect.
Figure~\ref{fig:jumpstat} shows the fraction of trajectories in the ensemble
that performed $n=0,1,2$ nonadiabatic transitions against time,
for the generalized sampling (main figure),
and the primitive sampling (inset).
Numerical fluctuations are reduced since the generalized weight 
makes ${\cal Q}=1$ when a transition is rejected.
Our calculations for the nonequilibrium dynamics
of the spin-boson model show that
the pruning allows one to sample longer times in a stable way,
while reproducing the short-time numerical results 
of the primitive algorithm, 
provided a suitable choice of $c_{\epsilon}$ is made.

These results, together with the calculations at lower friction
reported in~\cite{af}, provide substantial numerical evidence
that the generalized sampling is able to reduce the statistical error 
at long time.


\newpage

\newpage

\newpage

\newpage

\section{Discussion and conclusion}
\label{sec:c}

We have shown how our generalized sampling scheme 
can improve the numerical
stability of nonadiabatic dynamics at long times.
For devising our sampling scheme, we have been guided by 
the principle of energy conservation. In this paper, our choice of the
weight $w$ reflects such a line of thought. 
The results of our calculations illustrate in many relevant cases 
how big the reduction of the statistical error is at long time.
However, other possible choices of the weight $w$ (for example, better
suited or tailored to the problem at hand)
are open for investigation. The important idea is trying to get
a sequential weight equal to one (most of the times)
 when an adiabatic segment of
trajectory is  propagated. This limits the growth in time of
the statistical fluctuations induced by the sampling of nonadiabatic 
dynamics. One can speculate that other choices of $w$ can be tailored and
optimized for the specific process investigated.
We stress that these alternative choices of $w$
will all be encompassed by the general structure of
Eq.~(\ref{eq:prob-jump}).

We must remark that in order to reduce the systematic error
at high friction values, one is forced to use a higher-order approximation
of the short-time propagator, which has been first 
introduced in~\cite{trotter}. Here, our concern has been only
the reduction of the statistical error. To this end our sampling scheme
is very successful. As a matter of fact, the accurate calculations 
of~\cite{trotter} have been obtained with a simple filtering recipe,
which basically amounts to a redefinition of those values of the observable
which are too high as equal to a threshold constant~\cite{threshold}.
We believe that the systematic application of our computational theory 
can start an exciting time when the power of nonadiabatic algorithms,
stemming from the work of~\cite{kc}, can be reassessed and further
extended by combining our enhanced sampling scheme with higher order
approximations of the short-time propagator~\cite{trotter}.

It is also worth mentioning that our sampling scheme can be applied
to a more advanced version of Wigner-Heisenberg quantum(-classical) 
mechanics (which has been first proposed in~\cite{sk-wigner}
and further developed in~\cite{kim})
that can address time correlation functions
with quantum corrections to the structure of the bath coordinates
even with anharmonic potentials.



\section*{Acknowledgments}
This work is based upon research supported by the South
African Research Chair Initiative of the Department of
Science and Technology and National Research Foundation.


\end{document}